\documentclass[final]{svjour3}
\usepackage{graphicx}
\usepackage{rotating}
\usepackage{amssymb}
\usepackage{mathptmx}
\usepackage{subfigure}
\usepackage[usenames, dvipsnames]{color}
\usepackage{upgreek}
\usepackage{todonotes}
\usepackage{xcolor}
\makeatletter
\journalname{Journal of Low Temperature Physics}

\begin{document}

\newcommand{\hdblarrow}{H\makebox[0.9ex][l]{$\downdownarrows$}-}

\title{Reducing the susceptibility of lumped-element KIDs to two-level system effects}
\author{A.~L.~Hornsby\textsuperscript{a} \and P.~S.~Barry\textsuperscript{b,c} \and S.~M.~Doyle\textsuperscript{a} \and \\Q.~Y.~Tang\textsuperscript{b} \and E.~Shirokoff\textsuperscript{b}}
\institute{\textsuperscript{a} School of Physics and Astronomy, Cardiff University, Cardiff, The Parade, CF24 3AA, UK \\
\email{hornsbyal@cardiff.ac.uk}\\
 \textsuperscript{b}Department of Astronomy {{\&}} Astrophysics, University of Chicago, 5640 South Ellis Avenue, Chicago, IL 60637, USA\\
 \textsuperscript{c} Argonne National Laboratory, 9700 South Cass Avenue, Lemont, IL 60439, USA\\}
\maketitle
\begin{abstract}
Arrays of lumped-element kinetic inductance detectors (LEKIDs) optically coupled through an antenna-coupled transmission line are a promising candidate for future cosmic microwave background (CMB) experiments. However, the dielectric materials used for the microstrip architecture are known to degrade the performance of superconducting resonators. In this paper, we investigate the feasibility of microstrip coupling to a LEKID, focusing on a systematic study of the effect of depositing amorphous silicon-nitride on a LEKID. The discrete and spatially-separated inductive and capacitive regions of the LEKID allow us to vary the degree of dielectric coverage and determine the limitations of the microstrip coupling architecture. We show that by careful removal of dielectric from regions of high electric field in the capacitor, there is minimal degradation in dielectric loss tangent of a partially covered lumped-element resonator. We present the effects on the resonant frequency and noise power spectral density and, using the dark responsivity, provide an estimate for the resulting detector sensitivity.

\end{abstract}

\section{Introduction}
The temperature and polarisation anisotropies contained in the Cosmic Microwave Background (CMB) offer a unique window into how our Universe began. Detection of primordial B-modes is one of the main objectives for future CMB experiments and are critical to testing models of inflation. \cite{planck15EBmodes,GWinflation} Current experiments, with focal planes based on arrays of superconducting transition-edge sensors (TESs), now routinely operate with sensitivity close to the background-limit.  \cite{BK,SPT3G,ACTPOL} To further increase sensitivity, more detectors are required. It is now well known that observations at multiple frequencies are necessary in order to constrain and remove foreground contamination. \cite{BK} To maximise focal plane efficiency, each on-sky pixel includes dual-band, dual-polarisation sensitivity, with each pixel requiring four detectors. The number of detectors required by the next generation of CMB experiments presents a significant technical challenge.

Kinetic inductance detectors (KIDs) are superconducting resonators whose resonant frequency and quality factor are modified with absorbed optical power. \cite{KID_orig} Large arrays of KIDs can be constructed without the need for ancillary multiplexing components, significantly reducing the cryogenic complexity of an experiment. Currently, most existing CMB experiments implement an on-chip microstrip optical coupling architecture, where radiation is guided onto a thin-film microstrip line and then routed to the detector, enabling multi-chroic, polarisation-sensitive pixels. To take advantage of this heritage, and the separated inductive and capacitive elements of a lumped-element KID (LEKID), we have developed the microstrip-coupled LEKID (mc-LEKID).   \cite{LEKIDs,PBarry}

The goal of the mc-LEKID is to provide a simple and reliable design for microstrip coupling to a LEKID. Here, the resonator inductor doubles as both a high-Q microwave lumped-element inductor, as well as an absorbing mm-wave microstrip line (Fig \ref{schematic}, right). The microstrip line from the antenna is galvanically connected to the centre of a hairpin-style inductor. Radiation incident along the input microstrip line is split and is absorbed along the length of the inductor. Feeding the inductor at the voltage-null enables a direct connection to the microwave resonator without affecting the performance. However, the microstrip dielectric is formed from an amorphous material that is deposited over the KID, which has a tendency to degrade the performance of the KID resonator through the introduction of two-level systems (TLSs).  \cite{amor_si,amor_si2,tls_MUSIC}

In this paper, we investigate the effects on the microwave resonator caused by deposition of a silicon nitride (SiN\textsubscript{x}) dielectric layer over the LEKID architecture. In particular, we compare LEKID resonators with varying coverage of SiN\textsubscript{x} to study the impact on resonant frequency as a function of temperature and noise, as a first look at TLS contributions in a LEKID device architecture.

\section{Prototype device} 

\begin{figure} 
\begin{center}
    \includegraphics[width=0.82\linewidth, keepaspectratio,page=1]{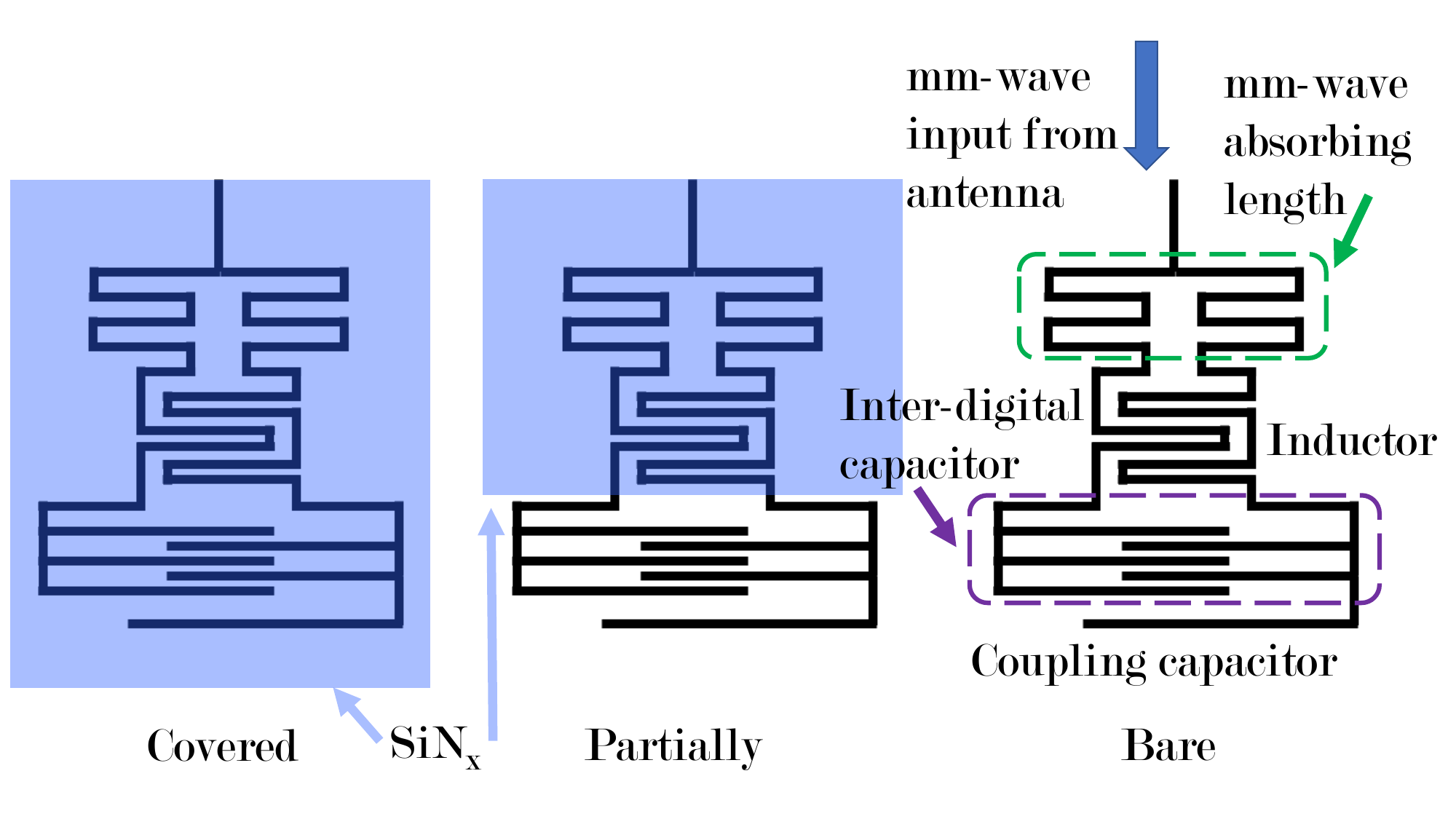}
	\caption{Schematic describing the three different dielectric scenarios presented in this paper. Left: SiN\textsubscript{x} covers the entire resonator. Centre: SiN\textsubscript{x} is over the inductor only. Right: Resonator free of dielectric. This addition of dielectric is necessary for transmission line coupling in antenna-coupled devices. (Colour figure online)}
\label{schematic}
\end{center}
\end{figure}

\noindent The prototype device presented here is a single-layer, dark array of 18 thin-film aluminium (Al) LEKIDs, with various SiN\textsubscript{x} coverage to investigate the impact of adding the dielectric to our detectors. Fig. \ref{schematic} shows a schematic version of the devices and highlights our three different scenarios. Here, our resonators are grouped into three frequency banks, which span (i) 786 - 801 MHz, (ii) 1029 - 1058 MHz and (iii) 1100 - 1166 MHz, each containing 6 resonators.

The lowest frequency resonators, bank (i), are completely covered by SiN\textsubscript{x} and are expected to have the highest loss (Fig. \ref{schematic}, left). The partially-covered resonators (Fig. \ref{schematic}, centre), bank (ii), have SiN\textsubscript{x} over the inductor only -- this is the desired dielectric coverage for the proposed coupling scheme. \cite{PBarry} The highest frequency bank of resonators, bank (iii), is completely free of dielectric (Fig. \ref{schematic}, centre). While this does not allow for optical coupling via a microstrip, this structure serves as a reference device for our proposed design.

During device fabrication, 50-nm layer of Al is first sputtered on to the Silicon wafer to ensure maximum control over processing the Al which, in turn,  reduces the number of TLSs created by the original substrate. \cite{Shirokoff} A wet-etchant is used to pattern the Al film. Next, 500 nm of SiN\textsubscript{x} is deposited in a high-density plasma chemical-vapour deposition system. To further reduce our susceptibility to TLSs, we fluorine-etch the SiN\textsubscript{x} from the different regions of the LEKID, depending on the desired level of dielectric coverage. For further details of the device fabrication process, see the Al and SiN\textsubscript{x} elements of Tang et al. \cite{Fab_paper}

\section{Results}

\begin{figure}[htbp]
\begin{center}
    \subfigure[]{\includegraphics[width=0.48\linewidth]{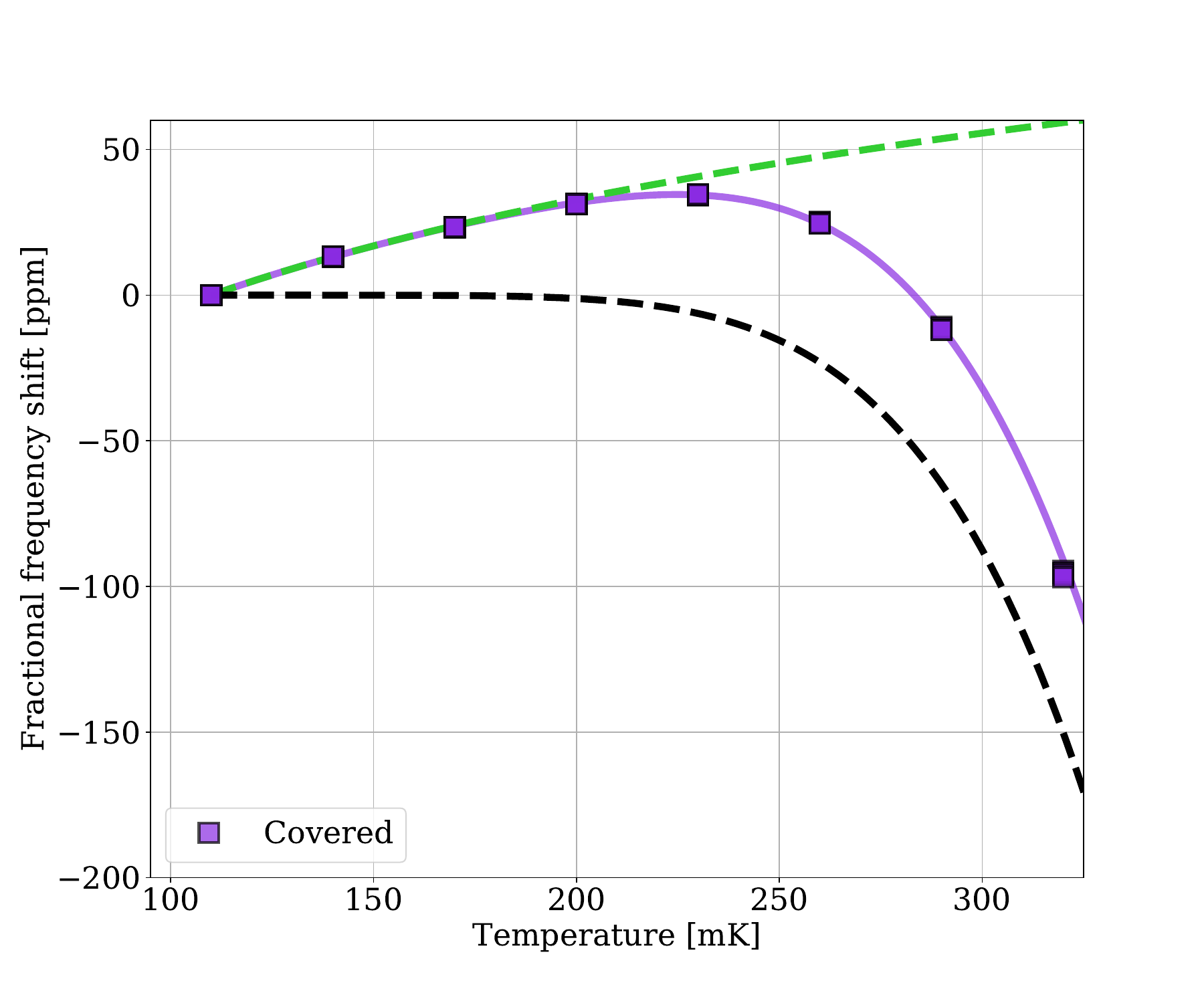}}
    \subfigure[]{\includegraphics[width=0.48\linewidth]{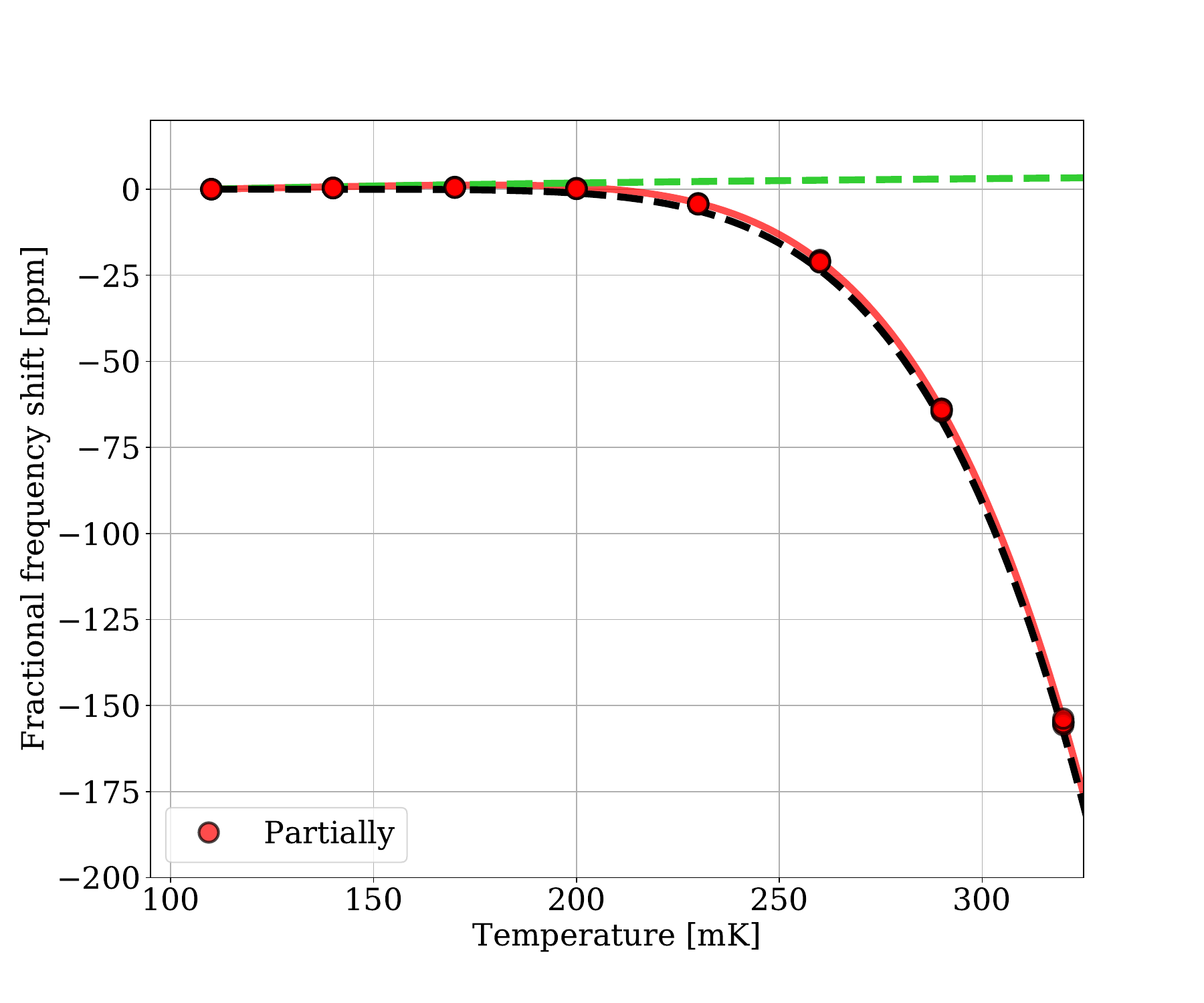}}
    \end{center}
\begin{center}
     \subfigure[]{\includegraphics[width=0.48\linewidth]{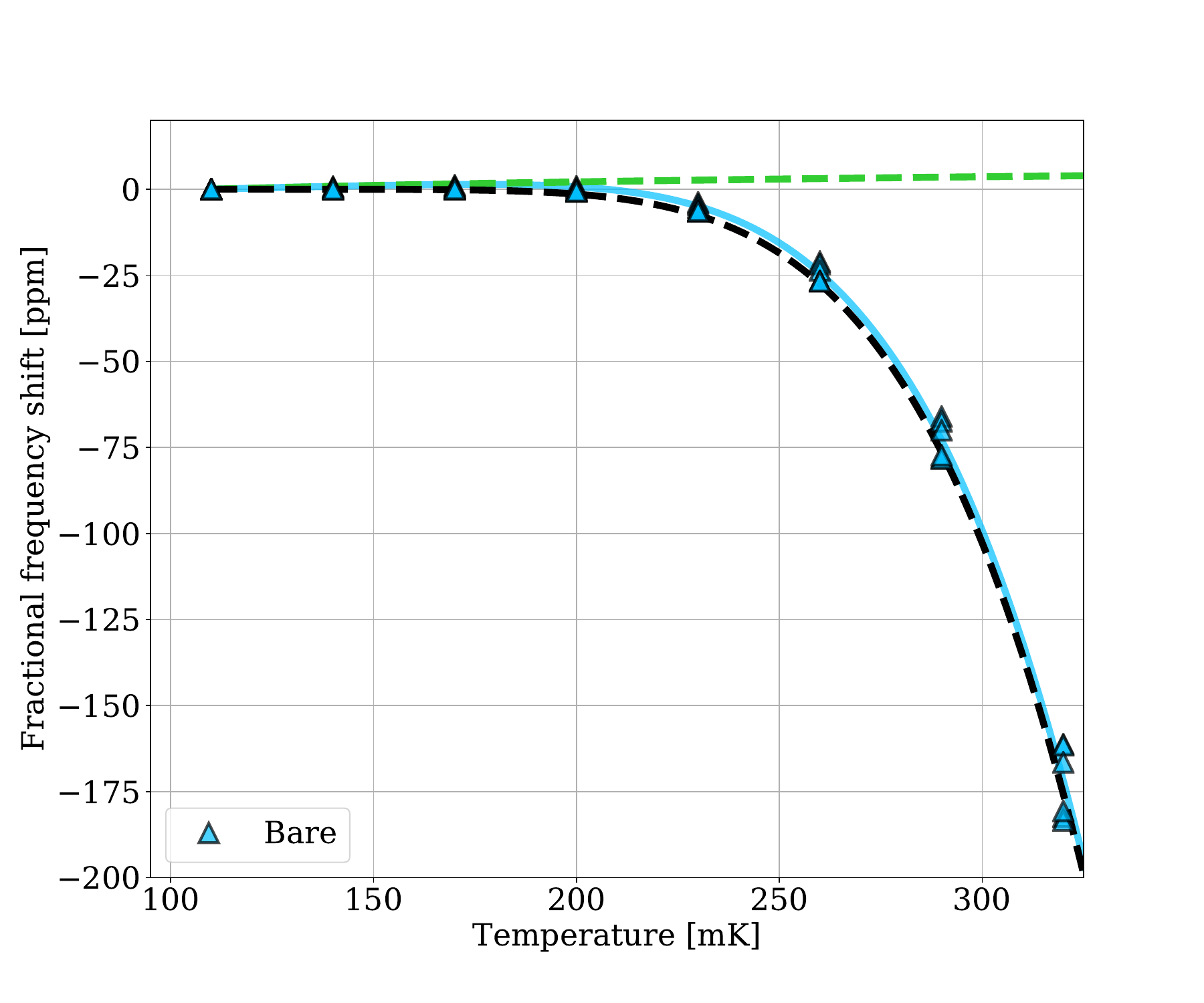}}
     \subfigure[]{\includegraphics[width=0.48\linewidth]{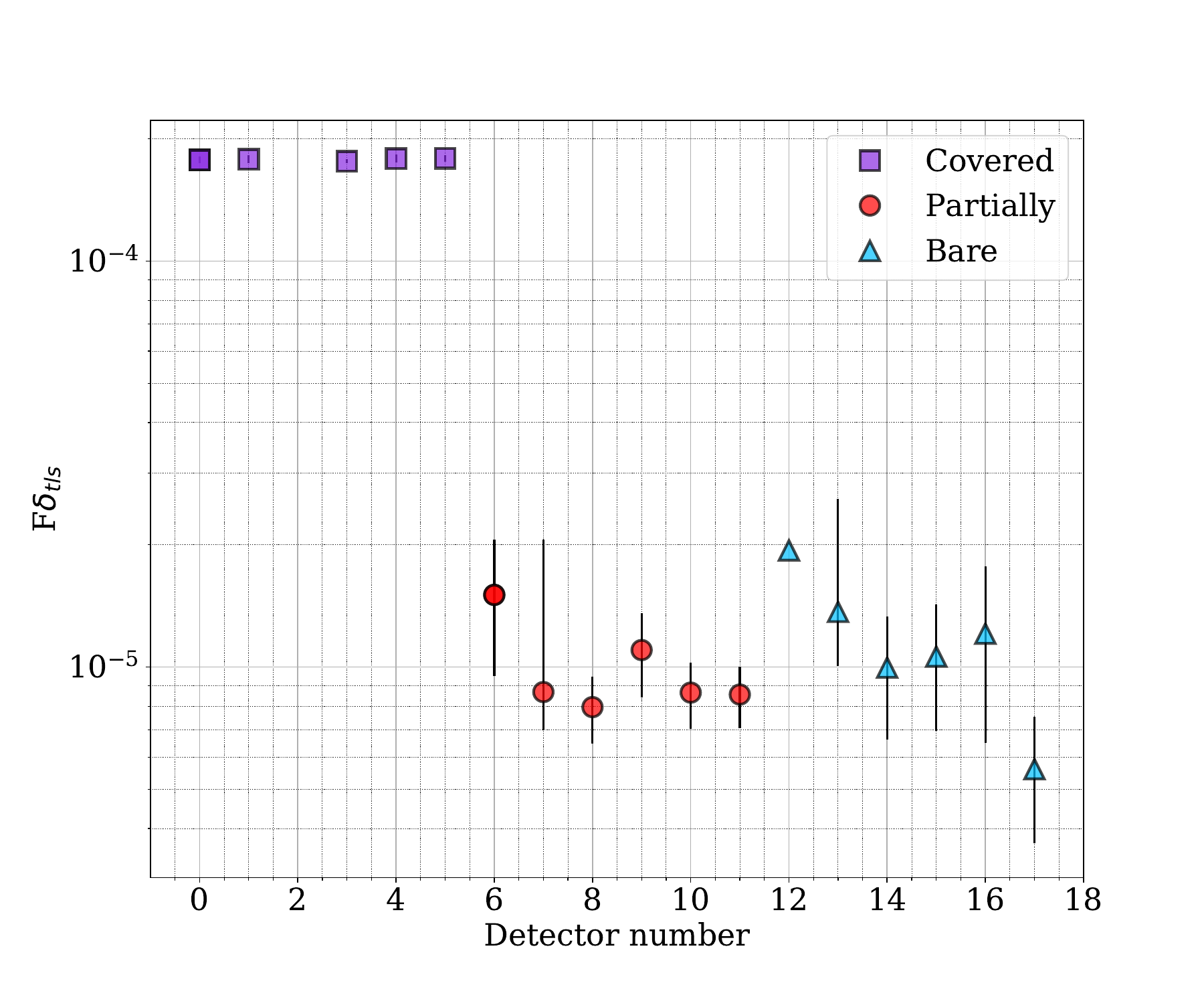}}

\caption{The fractional frequency shift as a function of temperature for our resonators, as described in Fig. \ref{schematic}, where (a) is covered, (b) is partially covered and (c) is completely free of by SiN$_{x}$. One covered resonator did not yield. The dashed lines are the two different components to the fit of the data - the expected fractional frequency shift (dashed \textit{black} line) and the additional shift caused by TLSs, calculated from equations \ref{eq:losses} and \ref{eq:tlsshift} (dashed \textit{green} line). The combined fit is a solid line. Plot (d) is the product F$\delta$\textsubscript{0} extracted from fitting the fractional frequency response of each resonator, where the error bars indicate the error in fit. (Colour figure online)}
\label{tls_fit}
\end{center}
\end{figure}

As a preliminary investigation into the impact of SiN\textsubscript{x} on the performance of the prototype device, we explored the dark response of the resonators outlined in Fig. \ref{schematic}. The prototype device sits in a gold-plated OFHC copper sample box installed on the 80 mK baseplate of a cryostat, cooled via a miniature dilution system. \cite{Teleberg} We measured S\textsubscript{21} of the resonators as a function of base temperature between 110 - 320 mK using a Vector Network Analyser (VNA), and extracted the resonant frequency following the procedure outlined in Khalil et al. \cite{resonator_fit} As determined by the bifurcation point, each detector was driven at their optimal readout power to reach a minimum detector noise and this varied between detectors. \cite{Swenson2013} Generally, partially-covered and covered resonators had a higher optimum drive power than bare resonators.

\subsection{Dark Response}

We investigated the fractional frequency shift of each resonator as a function of base temperature, as the first measure of detector performance. In Fig. \ref{tls_fit}(a)-(c), we observe the so-called back-bending in all devices, where the resonant frequency increases with temperature instead of decreasing, typical of TLSs perturbations. \cite{Dielectric_losses,barends2008,barends2010} Removing SiN\textsubscript{x} from the capacitive region of the LEKID reduces back-bending substantially, however there is no discernible difference between the partially-covered and bare resonator (Fig. \ref{tls_fit}(b) and (c)). Thus, the addition of SiN\textsubscript{x} to the LEKID architecture does not introduce a significant additional TLSs frequency response relative to the completely bare resonator. 

We can begin to quantify the effect of adding lossy dielectrics to our resonators by considering the product of the fill factor and the TLS-induced loss tangent $F\delta_{0}$. \cite{Dielectric_losses,barends2008,barends2010}
\begin{equation} \frac{\upDelta f_{r}}{f_{r}} = - \frac{F}{2} \frac{\upDelta \epsilon}{\epsilon},
\label{eq:losses}
\end{equation} 
where $\epsilon$ is the dielectric constant parametrised as,  
\begin{equation}
\frac{\upDelta \epsilon}{\epsilon} = \frac{2\delta_0}{\pi} \Bigg[ \textrm{Re} \Psi \Big( \frac{1}{2} + \frac{1}{2 \pi i}\frac{\hbar \omega}{kT}\Big) - \log_e \frac{\hbar \omega}{kT} \Bigg],
\label{eq:tlsshift}
\end{equation}

\noindent $\omega$ is the frequency, $\Psi$ is the complex digamma function. Here, we fit a combined model of the additional frequency response from the dielectric and the expected response from Mattis-Bardeen superconductivity. From the fits (solid lines in Fig. \ref{tls_fit}(a)-(c)), we found removing the dielectric from the capacitive region of the LEKID, on average, reduced $F\delta_{0}$ from (1.78$\pm$0.02) x10$^{-4}$ to (1.03$\pm$0.11) x10$^{-5}$. On average, this equates to a reduction in $F\delta_{0}$ of $\sim$ 16 across the partially-covered and bare resonators of the prototype array. The average $F\delta_{0}$ was found to be (1.14$\pm$0.18) x10$^{-5}$ for the bare resonators.  It should be noted that the capacitor geometry for each LEKID is very similar, meaning F should only vary slightly between detectors meaning this experiment provides good insight into the dielectric loss tangent. 

\subsection{Dark Detector Noise}
To characterise the noise of the detectors, and estimate the electrical noise equivalent power (NEP), we utilised the standard single-tone homodyne readout. \cite{KID_orig}

\begin{figure}[htbp]
\begin{center}
    \subfigure[]{\includegraphics[width=0.48\linewidth]{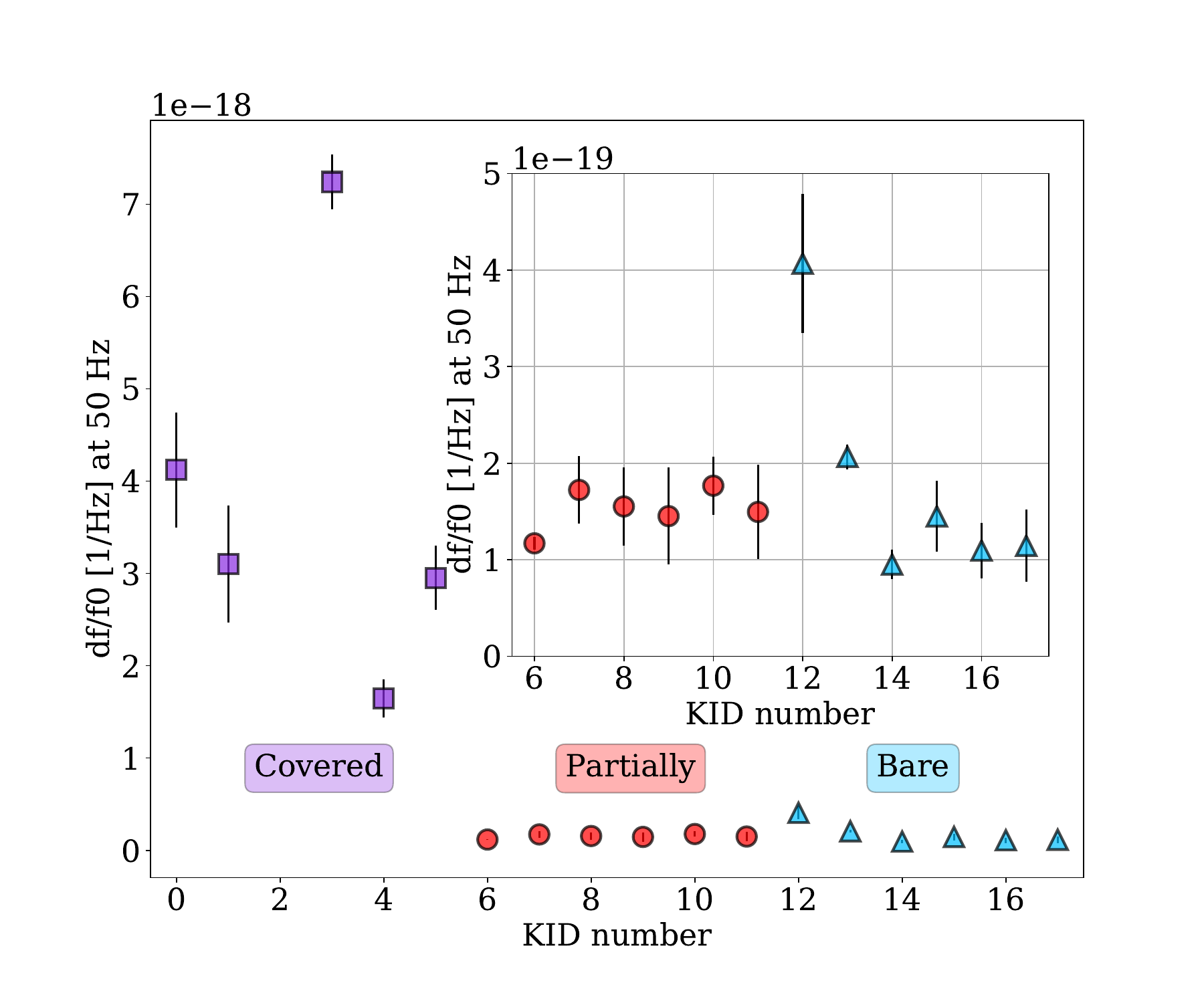}}
    \subfigure[]{\includegraphics[width=0.48\linewidth]{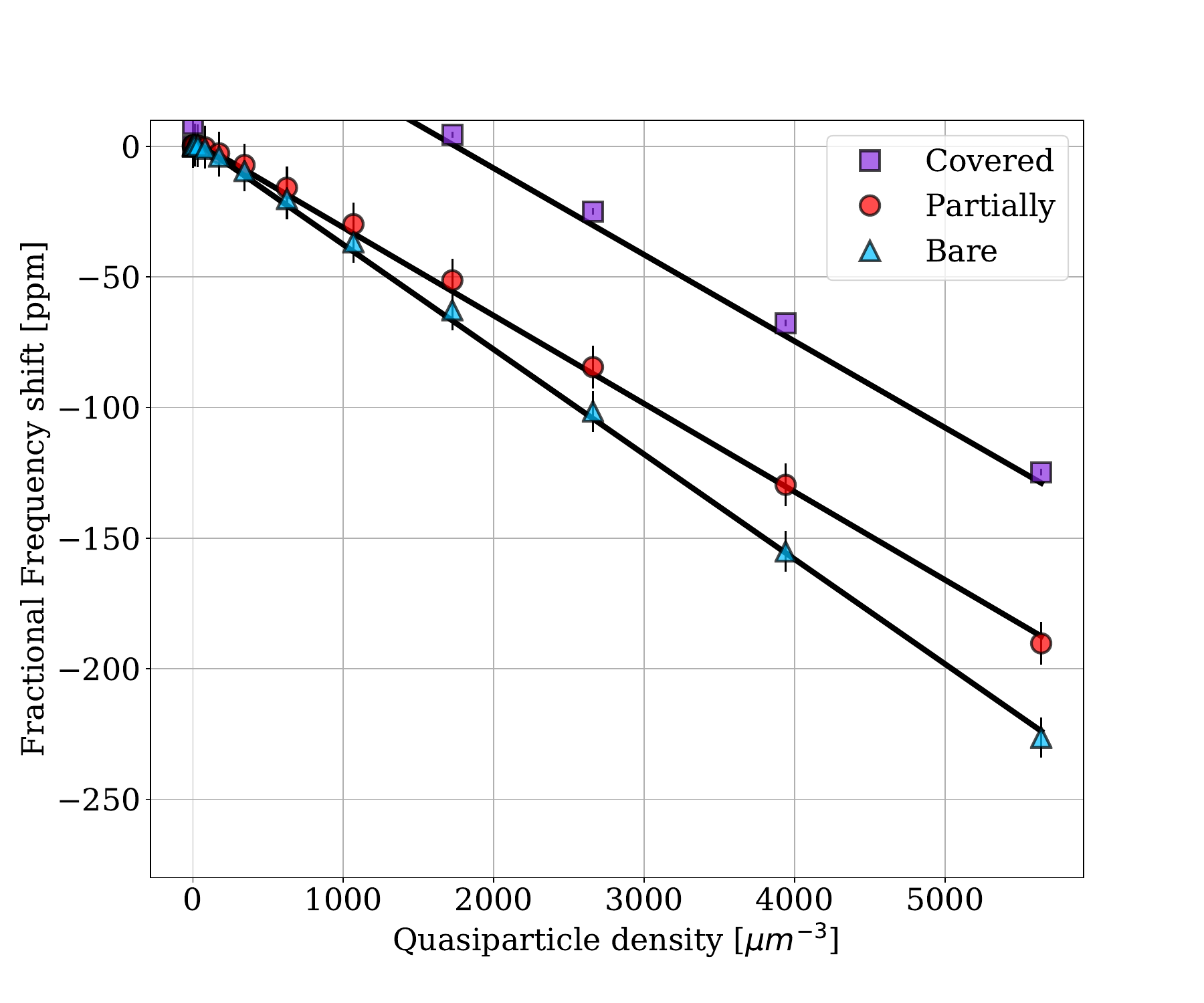}}
\end{center}
\begin{center}
    \subfigure[]{\includegraphics[width=0.48\linewidth]{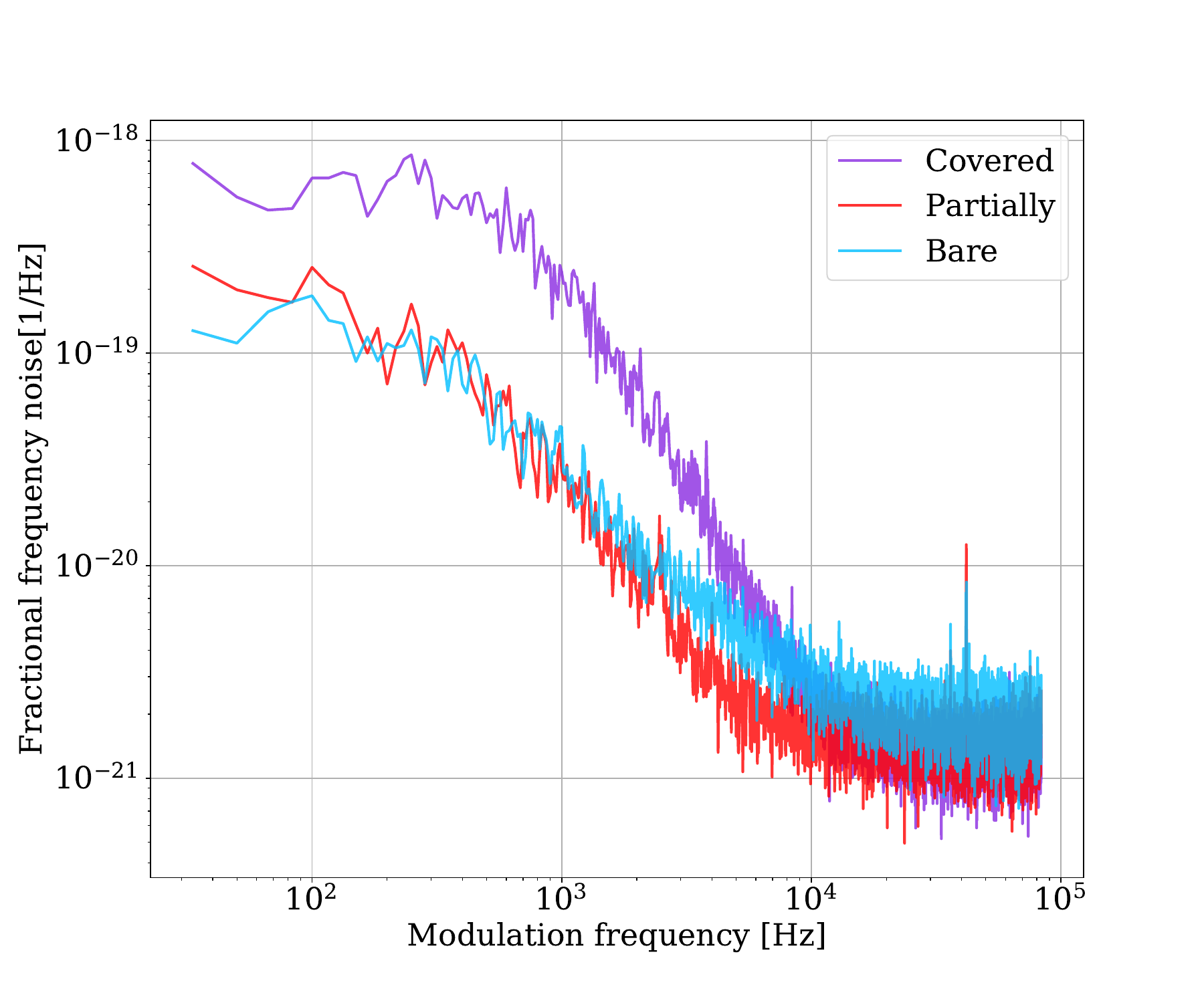}}
    \subfigure[]{\includegraphics[width=0.48\linewidth]{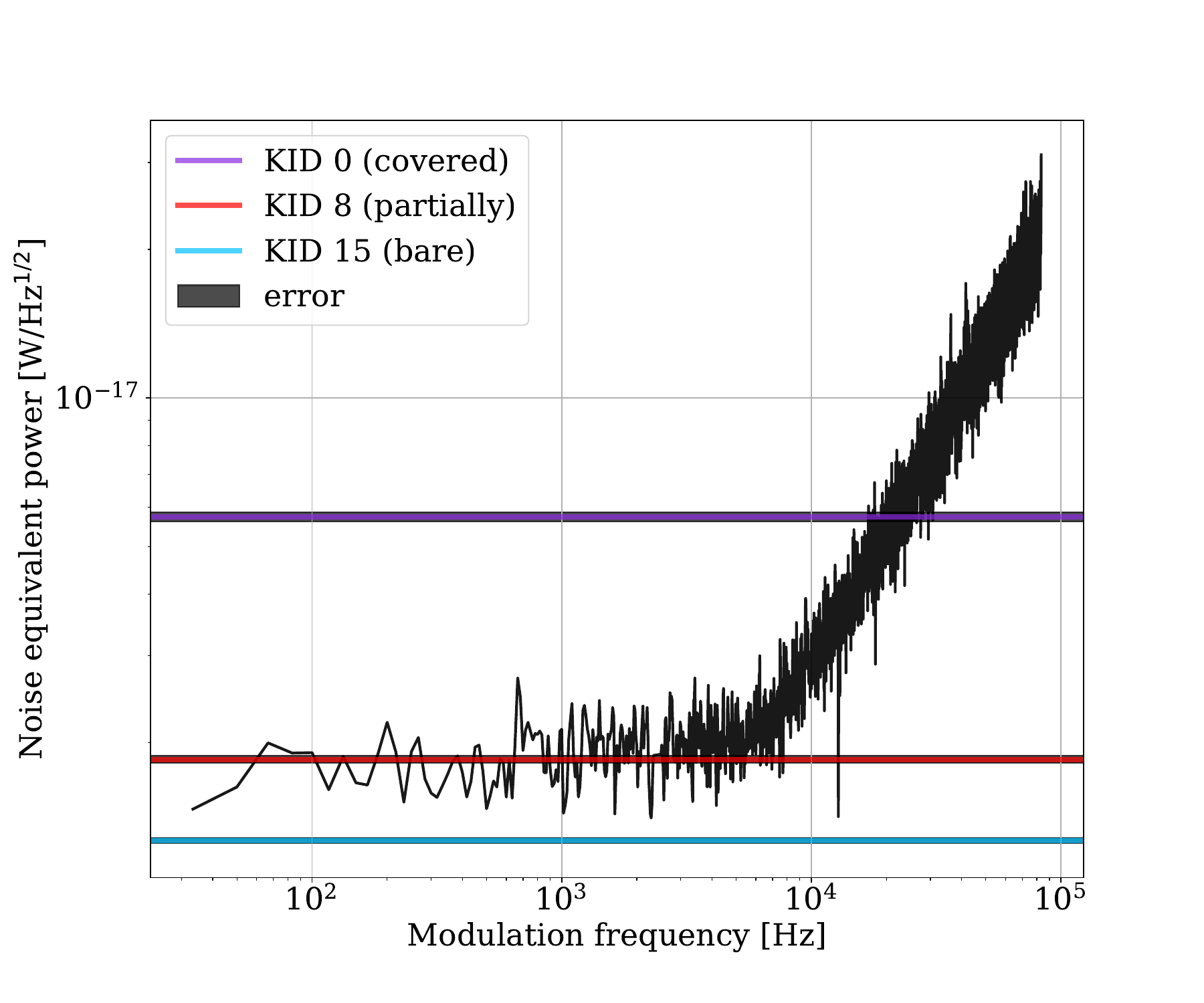}}
    \caption{(a) Measuring the fractional frequency noise at a modulation frequency of 50 Hz for covered (\textit{purple}), partially-covered (\textit{red}) and bare (\textit{blue}) resonators at 80 mK. One covered resonator did not yield. The inset in (a) allows for a closer look at df/f\textsubscript{0} for the bare and partially-covered resonators. An average of several data points was used to determine the fractional frequency noise level at 50 Hz, thus the error here represents the range of results (b) Estimating the responsivity of a covered (\textit{purple}), partially-covered (\textit{red}) and bare (\textit{blue}) KID from the linear response of the resonator to the changing quasi-particle density as a function of base temperature. Only higher temperature data points are fitted due to the presence of back-bending, and the error shown is the extracted error in fit (c) The measured noise spectrum of a covered (\textit{purple}), partially-covered (\textit{red}) and bare (\textit{blue}) resonator at 160 mK. (d) Dark noise equivalent power calculated (\textit{black}) after the noise power spectral density of a partially-covered resonator was corrected for the quasi-particle roll off. The measured dark NEP values for the the covered (\textit{purple}), partially-covered (\textit{red}) and bare (\textit{blue}) are also shown here. (Colour figure online)}
\label{noise_a_b}
\end{center}
\end{figure}
In Fig. \ref{noise_a_b}(a), we show the fractional frequency noise (df/f\textsubscript{0}) of each detector at a modulation frequency of 50 Hz to compare detector noise in the flat region of the noise power spectral density. Overall, we found the covered resonators had the highest noise level of all resonators with an average of (3.81$\pm$0.20) x10$^{-18}$ Hz$^{-1}$, compared to the partially-covered at (1.53$\pm$0.16) x10$^{-19}$ Hz$^{-1}$ and the bare resonators at (1.80$\pm$0.17) x10$^{-19}$ Hz$^{-1}$. Thus, the average noise power level is $\sim$ 25 times higher compared to the partially-covered resonator and $\sim$ 21 times higher than the bare resonator, which indicates adding the dielectric to the absorbing element of the detector does not result in excess noise being observed. Moreover, this suggests that noisier dielectric materials could be used with this detector architecture, but this requires testing of additional devices in the future.

We estimate the responsivity of resonators 0, 8 and 15 by fitting the fractional frequency response as a function of quasi-particle number (see Fig. \ref{noise_a_b}(b)). Low temperature data points are excluded due to back-bending and the error shown here is the error in fit. With the fractional frequency noise level and detector responsivity calculated from the noise power spectral density (see Fig. \ref{noise_a_b}(c)), alongside known material parameters, we can determine the dark NEP from Baselmans et al. \cite{Baselmans} The resulting NEP, derived from the responsivity and corrected for quasi-particle roll off, is shown in Fig. \ref{noise_a_b}(d) for the partially-covered resonator 8, alongside the NEP calculated for resonators 0 and 15. The NEP values are (5.73$\pm$0.12) x10$^{-18}$ W Hz$^{-1/2}$ for the covered resonator (KID 0), (1.85$\pm$0.03) x10$^{-18}$ W Hz$^{-1/2}$ for the partially-covered resonator (KID 8) and (1.28$\pm$0.02) x10$^{-18}$ W Hz$^{-1/2}$ for the bare resonator (KID 15).

To fully characterise the detector noise properties we need to measure the 1/f knee of the noise spectrum. However, excessive system 1/f noise is currently prohibiting low-frequency measurements. Going forward, the noise of each detector must be measured simultaneously to allow for de-correlated noise analysis and to enable the exploration of low-frequency behaviour.

\section{Conclusion}
Optical coupling to a LEKID via an antenna and transmission line structure is a promising candidate for future CMB experiments requiring large detector arrays, making possible the addition of structures needed for multi-chroic, polarisation-sensitive capabilities.  Separating the inductive and capacitive elements allows for additional TLS losses, caused by placing dielectric over the resonator, to be minimised. For a dark, single-layer LEKID array, we have demonstrated that we can meet the dielectric requirements of the microstrip transmission line coupling, whilst maintaining a minimal additional parasitic dielectric response and without creating additional noise.

\begin{acknowledgements} 
We acknowledge support from the Science and Technology Facilities Council (STFC) Consolidated grant Ref: ST/N000706/1. This work is also supported by the National Science Foundation under Grant Number 1554565 and made use of the Pritzker Nanofabrication Facility of the Institute for Molecular Engineering at the University of Chicago, which receives support from SHyNE, a node of the National Science Foundation’s National Nanotechnology Coordinated Infrastructure (NSF NNCI-1542205). 
\end{acknowledgements}

\end{document}